# Variance estimation for logistic regression in case-cohort studies

Hisashi Noma, PhD[*]

Department of Data Science, The Institute of Statistical Mathematics, Tokyo, Japan

ORCID: http://orcid.org/0000-0002-2520-9949

*Corresponding author: Hisashi Noma

Department of Data Science, The Institute of Statistical Mathematics

10-3 Midori-cho, Tachikawa, Tokyo 190-8562, Japan

TEL: +81-50-5533-8440

e-mail: noma@ism.ac.jp

**Abstract**

**Background:** The logistic regression analysis proposed by Schouten et al. (Stat Med. 1993;12:1733–1745) has been a standard method in current statistical analysis of case-cohort studies, and it enables effective estimation of risk ratio from selected subsamples, with adjustment of potential confounding factors. Schouten et al. (1993) also proposed the standard error estimate of the risk ratio estimator can be calculated by the robust variance estimator, and this method has been widely adopted.

**Methods and Results:** We show that the robust variance estimator does not account for the duplications of case and subcohort samples and generally has certain bias, i.e., inaccurate confidence intervals and P-values are possibly obtained. To address the invalid statistical inference problem, we provide an alternative bootstrap-based valid variance estimator. Through simulation studies, the bootstrap method consistently provided more precise confidence intervals compared with those provided by the robust variance method, while retaining adequate coverage probabilities.

**Conclusion:** The robust variance estimator has certain bias, and inadequate conclusions might be deduced from the resultant statistical analyses. The proposed bootstrap variance estimator can provide more accurate and precise interval estimates. The bootstrap method would be an alternative effective approach in practice to provide accurate evidence.

Key words: case-cohort design, logistic regression, risk ratio, bias, bootstrap.

**Introduction**

The case-cohort design [1] has been widely used as an efficient study design to reduce costs and effort for clinical and epidemiologic studies. In statistical analyses of case-cohort studies, the logistic regression analysis proposed by Schouten et al. [2] has been one of the standard methods used in current practice, and it enables risk ratios to be effectively estimated from the selected subsamples. A remarkable advantage of their method is that the computation can be easily implemented using statistical packages for ordinary logistic regression analysis (e.g., `glm` in R). In the logistic regression analysis, duplicate participants between case and subcohort samples are formally regarded as different participants and a logistic regression is fitted to the selected subsamples as ordinary case-control studies [3]:

$$\text{logit}\{\Pr(D = 1)\} = \beta_0 + \beta_1 x_1 + \cdots + \beta_p x_p$$

where $D$ is an indicator variable that is equal to 1 if a participant is in the case samples and 0 if he/she is in the subcohort samples. Also, $x_1, \ldots, x_p$ are the explanatory variables. The formal maximum likelihood estimators of the regression coefficients except for the intercept $\beta_0$ have been shown to be unbiased (consistent) estimators of the log risk ratios in the target population [2]. Schouten et al. [2] noted that their standard errors can be estimated by the robust (sandwich) variance estimator. However, this variance estimation ignores the duplications of case and subcohort samples and simply fits the ordinary robust variance formulae to the pseudo-likelihood function. In the present article, we show that the variance estimator is biased because the duplications are not adequately accounted for and that the resultant confidence intervals are usually unprecise. We also provide an alternative valid variance estimator using bootstrap and show its effectiveness via simulation studies.

**Methods**

*Bias of the robust variance estimator*

Consider a cohort that consists of $N$ participants. We here consider a case-cohort study that samples $n_1$ case samples among the cases and $n_0$ subcohort samples among the entire cohort. Schouten et al. [2] proposed that the risk ratio of the target population is consistently estimated by the ordinary logistic regression and that the standard errors of the regression coefficient are also consistently estimated by the robust variance estimator, because they considered $D$ is correlated for the duplicated samples. However, they did not confirm the validity of their approach via rigorous asymptotic theory or simulation studies. $D$ is actually not correlated for the duplicated samples, because the subcohort sampling is independently determined by the case status. Thus, the model and robust variance estimators are asymptotically equivalent. Also, from the asymptotic theory of response-selective design [4], especially for the conventional case-control study [3, 5], the consistency of the Prentice-Pyke-type model variance estimator [3] is assured when the all case and subcohort samples are unduplicated. However, duplications usually exist between the case and subcohort samples and the assumption is generally violated. Thus, the robust variance estimator is biased because the duplications are not adequately accounted for. In subsequent simulation studies, the robust variance estimator certainly overestimated the actual standard errors of the regression coefficients (Table 1).

*Alternative valid variance estimator using bootstrap*

To construct unbiased variance estimators, the duplications should be adequately accounted for; consider that $m$ duplicate samples exist between the case and subcohort samples. An effective approach is to use the bootstrap method. To account for the duplications, we formally consider the sampling scheme of a case-cohort study as a conventional case-control sampling; Noma and Tanaka [6] showed that the sampling design of case-cohort studies is theoretically equivalent to that of case-control sampling. The duplicated samples are then regarded as randomly selected samples from the case samples. To quantify the uncertainty adequately, an effective approach is to incorporate this sampling mechanism by bootstrap. The bootstrap algorithm is then given as follows.

*Algorithm (bootstrap variance estimation)*

1. Perform a bootstrap resampling from the $n_1$ case samples.
2. Perform a bootstrap resampling from the $n_0 - m$ non-case samples in the subcohort.
3. Select $m$ samples from the $n_1$ bootstrap samples of case samples randomly, and add them to the subcohort (the duplicated samples).
4. Fit the logistic regression to the bootstrap samples generated by processes 1–3.
5. Repeat processes 1–4 and calculate bootstrap samples of the regression coefficient estimates sufficient times. Then, compute the empirical variances of the bootstrap samples of regression coefficients.

Through the bootstrap algorithm, consistent standard error estimates can be obtained. In processes 2 and 3, the duplications are adequately accounted for in the bootstrap algorithm. An alternative asymptotically equivalent resampling strategy is to resample from the $n_1 - m$ unduplicated case samples, the $n_0 - m$ non-case samples, and the $m$ duplicated samples, separately. Also, we can consider another naïve bootstrap strategy that substitutes processes 2 and 3 with process 2′:

2′. Perform a bootstrap resampling from the $n_0$ samples in the subcohort.

This bootstrap resampling corresponds to the naïve bootstrap for ordinary case-control studies, and the duplications are not accounted for. This naïve bootstrap algorithm provides standard error estimates similar to those of Schouten et al.'s [2] robust variance estimator.

An R package `bootcc` (https://github.com/nomahi/bootcc) is available for implementing the proposed bootstrap inference method by simple commands.

*Simulation studies*

To assess the validities of the theoretical results, we carried out simulation studies. The simulation settings were based on the Wilms' tumor studies of Breslow et al. [7] For the event occurrence mechanisms, we considered a binomial regression model with a log link function, $\log\{\Pr(Y = 1 \mid z, x_1, x_1)\} = \beta_0 + \beta_1 z + \beta_2 x_1 + \beta_3 x_2$, where $x_2$ and $x_3$ are dummy variables of a trinomial distribution with event probabilities 0.16 and 0.48, respectively, and $x_1$ is a binary variable that was determined by a logistic regression model, $\text{logit}\{\Pr(Z = 1 \mid r_1, r_2)\} = \gamma_0 + \gamma_1 r_1 + \gamma_2 r_2$, where $r_1 \sim$ Bernoulli (0.10) and $r_2 \sim N(0, 1)$. Variable $Z$ is considered an expensive covariate measured only for the subsamples selected in the case-cohort design. The expected event fraction was 15.4%. The sample size of entire cohort $N$ was set to 2000, 4000, and 10,000, and the subcohort size was determined to be 20% and 40% of $N$. Also, all cases were sampled as case samples. The number of bootstrap resampling was consistently set to 2000, and 10,000 simulations were performed for all scenarios.

**Results**

The simulation results are presented in Table 1. We assessed the means and standard deviations of the regression coefficient estimates across the 10,000 simulations. We also evaluated the means of standard error estimates across the 10,000 simulations for the robust variance estimator ($\widehat{SE}_{robust}$), the naïve bootstrap variance estimator ($\widehat{SE}_{boot,naive}$), and the proposed bootstrap variance estimator ($\widehat{SE}_{boot,proposed}$). In addition, we evaluated the empirical coverage probabilities of Wald-type 95% confidence intervals of the regression coefficients based on the three variance estimators (i.e., $CP_{robust}$, $CP_{boot,naive}$, and $CP_{boot,proposed}$).

The regression coefficients were unbiasedly estimated using the logistic regression analysis method of Schouten et al. [2] In addition, the means of the standard error estimates obtained by the robust variance estimators were certainly biased from the actual standard errors for all scenarios and overestimation biases were indicated. Furthermore, the means of standard error estimates obtained by the robust variance and the naïve bootstrap variance estimators were similar under all settings. These results indicate that the robust variance estimator did not account for the duplications of the samples. In addition, the proposed bootstrap variance estimators unbiasedly estimated the actual standard errors under all settings. The coverage rates of the 95% confidence intervals reflected these properties, and the confidence intervals obtained by the robust variance and the naïve bootstrap variance estimators were generally too conservative. More precise and valid interval estimates were provided by the proposed bootstrap method, with retention of adequate coverage rates.

## Discussion and Conclusions

Logistic regression analysis has been a standard method for case-cohort studies, and the robust variance estimator has been used for these analyses. As shown in the present study, the robust variance estimator has certain bias, and inadequate conclusions might be deduced from the resultant statistical analyses. By contrast, the proposed bootstrap variance estimator is shown to be unbiased. The resultant confidence intervals are more precise, and more accurate interval estimates are generally obtained. Thus, the bootstrap method should be adopted in practice to provide accurate evidence.

Recently, alternative efficient inverse probability weighting methods have been established for case-cohort studies [6,8]. However, because of its simplicity and usefulness, the logistic regression analysis will continue to be used as a standard method for case-cohort studies. Our results enable more accurate and precise evaluations of effect measures in case-cohort studies and would facilitate the use of Schouten et al.'s [2] effective method in practice.

# References

1. Prentice RL. A case-cohort design for epidemiologic cohort studies and disease prevention trials. *Biometrika* 1986; **73**: 1-11.

2. Schouten EG, Dekker JM, Kok DF, et al. Risk ratio and rate ratio estimation in case-cohort designs: Hypertension and cardiovascular mortality. *Stat Med* 1993; **12**: 1733-45.

3. Prentice RL, Pyke R. Logistic disease incidence models and case-control studies. *Biometrika* 1979; **66**: 403-11.

4. Lawless JF, Kalbfleisch JD, Wild CJ. Semiparametric methods for response-selective and missing data problems. *J Royal Stat Soc B* 1999; **61**: 413-38.

5. Breslow NE, Robins JM, Wellner JA. On the semi-parametric efficiency of logistic regression under case-control sampling. *Bernoulli* 2000; **6**: 447-55.

6. Noma H, Tanaka S. Analysis of case-cohort designs with binary outcomes: Improving the efficiency using whole cohort auxiliary information. *Stat Methods Med Res* 2017;

   **26**: 691-706.

7. Breslow NE, Chatterjee N. Design and analysis of two-phase studies with binary outcome applied to Wilms Tumour Prognosis. *J Royal Stat Soc C* 1999; **48**: 457-68.

8. Noma H, Misumi M, Tanaka S. Risk ratio and risk difference estimation in case-cohort studies. *J Epidemiol*; DOI: 10.2188/jea.JE20210509.

Table 1. Results of the simulations for the logistic regression analysis *.

| $N$ | 2000 | | 4000 | | 10,000 | |
|---|---|---|---|---|---|---|
| Subcohort size | 20% | 40% | 20% | 40% | 20% | 40% |
| Mean duplicated samples | 61.4 | 122.6 | 122.8 | 245.5 | 307.1 | 614.6 |
| $\beta_1 = 0.96$ | | | | | | |
| Mean | 0.969 | 0.962 | 0.965 | 0.963 | 0.962 | 0.960 |
| SE | 0.193 | 0.149 | 0.134 | 0.104 | 0.084 | 0.066 |
| $\widehat{SE}_{robust}$ | 0.217 | 0.182 | 0.153 | 0.128 | 0.096 | 0.081 |
| $\widehat{SE}_{boot,naive}$ | 0.221 | 0.184 | 0.154 | 0.129 | 0.097 | 0.081 |
| $\widehat{SE}_{boot,proposed}$ | 0.191 | 0.148 | 0.133 | 0.103 | 0.083 | 0.065 |
| $CP_{robust}$ | 0.974 | 0.983 | 0.975 | 0.985 | 0.978 | 0.986 |
| $CP_{boot,naive}$ | 0.976 | 0.985 | 0.976 | 0.986 | 0.979 | 0.986 |
| $CP_{boot,proposed}$ | 0.949 | 0.946 | 0.946 | 0.947 | 0.949 | 0.949 |
| $\beta_2 = -0.28$ | | | | | | |
| Mean | −0.280 | −0.286 | −0.283 | −0.282 | −0.281 | −0.281 |
| SE | 0.211 | 0.182 | 0.147 | 0.127 | 0.094 | 0.080 |
| $\widehat{SE}_{robust}$ | 0.232 | 0.205 | 0.164 | 0.144 | 0.103 | 0.091 |
| $\widehat{SE}_{boot,naive}$ | 0.235 | 0.208 | 0.165 | 0.145 | 0.103 | 0.091 |
| $\widehat{SE}_{boot,proposed}$ | 0.214 | 0.182 | 0.150 | 0.128 | 0.094 | 0.080 |
| $CP_{robust}$ | 0.970 | 0.976 | 0.973 | 0.975 | 0.969 | 0.975 |
| $CP_{boot,naive}$ | 0.972 | 0.978 | 0.973 | 0.976 | 0.969 | 0.975 |
| $CP_{boot,proposed}$ | 0.955 | 0.954 | 0.953 | 0.954 | 0.949 | 0.951 |
| $\beta_3 = -0.39$ | | | | | | |
| Mean | −0.392 | −0.391 | −0.390 | −0.390 | −0.391 | −0.390 |
| SE | 0.156 | 0.131 | 0.109 | 0.092 | 0.069 | 0.059 |
| $\widehat{SE}_{robust}$ | 0.169 | 0.149 | 0.119 | 0.105 | 0.075 | 0.066 |
| $\widehat{SE}_{boot,naive}$ | 0.170 | 0.150 | 0.120 | 0.105 | 0.075 | 0.067 |
| $\widehat{SE}_{boot,proposed}$ | 0.154 | 0.131 | 0.108 | 0.092 | 0.068 | 0.058 |
| $CP_{robust}$ | 0.966 | 0.976 | 0.969 | 0.976 | 0.968 | 0.973 |
| $CP_{boot,naive}$ | 0.967 | 0.977 | 0.969 | 0.976 | 0.968 | 0.972 |
| $CP_{boot,proposed}$ | 0.948 | 0.950 | 0.948 | 0.948 | 0.948 | 0.949 |

* *Mean, SE: Mean and SD of the regression coefficients estimates. $\widehat{SE}_{robust}, \widehat{SE}_{boot,naive}, \widehat{SE}_{boot,proposed}$: Means of the standard error estimates by robust and bootstrap variance estimators. $CP_{robust}, CP_{boot,naive}, CP_{boot,proposed}$: Coverage probabilities of Wald-type 95% confidence intervals by robust and bootstrap variance estimators.*